\documentclass[a4paper, oneside, twocolumn, notitlepage, 10pt]{extarticle_ecoc}
\usepackage[mode=buildmissing]{standalone}
\usepackage{ecoc}
\usepackage{upgreek}
\usepackage{kitcolors}
\usepackage{amsmath}
\usepackage{mleftright}
\DeclareMathOperator*{\argmax}{arg\,max}
\DeclareMathOperator*{\argmin}{arg\,min}
\usepackage{mathtools}
\usepackage{hyperref}
\usepackage{siunitx}
\usepackage{tikz}
\usepackage{bm}
\usepackage{algorithm}
\usepackage{algpseudocode}
\usepackage{balance}
\DeclareSIUnit{\belmilliwatt}{Bm}
\DeclareSIUnit{\dBm}{\deci\belmilliwatt}
\algnewcommand{\LineComment}[1]{\State \(\triangleright\) #1}

\newcommand\extrafootertext[1]{%
    \bgroup
    \renewcommand\thefootnote{\fnsymbol{footnote}}%
    \renewcommand\thempfootnote{\fnsymbol{mpfootnote}}%
    \footnotetext[0]{#1}%
    \egroup
}

\begin{document}
\selectlanguage{american}

%-------------------------------------------------- Title -----------------------------------------------------%

\title{Approximate Maximum a Posteriori Carrier Phase Estimator for Wiener Phase Noise Channels using Belief Propagation}%

%------------------------------------------------- Authors-----------------------------------------------------%

\author{Shrinivas Chimmalgi, Andrej Rode, Luca Schmid, Laurent Schmalen  
}

\maketitle                  % Create title and author

%------------------------------------------ Description of Authors ----------------------------------------------%

\begin{strip}
 \begin{author_descr}

   Communications Engineering Lab (CEL), Karlsruhe Institute of Technology (KIT), \textcolor{blue}{\uline{s.chimmalgi@kit.edu}}

   % \textsuperscript{(2)} Authors' full affiliation,
   % \textcolor{blue}{\uline{author@institution.org}} (Email address optional)

   % \textsuperscript{(3)} Authors' full affiliation,
   % \textcolor{blue}{\uline{author@institution.org}} (Email address optional)

 \end{author_descr}
\end{strip}

\setstretch{1.1}
%-------------------------------------------------- Footnote -------------------------------------------------------%
\renewcommand\footnotemark{}
\renewcommand\footnoterule{}
%\let\thefootnote\relax\footnotetext{text}

%-------------------------------------------------- Abstract ---------------------------------------------------------%

\begin{strip}
  \begin{ecoc_abstract}
    % NOTE: Don't use a blank line here but start abstract right away to avoid an extra line break
    The blind phase search (BPS) algorithm for carrier phase estimation is known to have sub-optimal performance for probabilistically shaped constellations. We present a belief propagation based approximate maximum a posteriori carrier phase estimator and compare its performance with the standard and an improved BPS algorithm. \textcopyright2023 The Author(s)
  \end{ecoc_abstract}
\end{strip}

%-------------------------------------------------- Introduction Section -------------------------------------------------------%

\section{Introduction}
\extrafootertext{This work has received funding from the European Research Council (ERC) under the European Union's Horizon 2020 research and innovation programme (grant agreement No. 101001899).}\extrafootertext{This paper is a preprint of a paper submitted to ECOC 2023 and is subject to Institution of Engineering and Technology Copyright. If accepted, the copy of record will be available at IET Digital Library.}
Probabilistic amplitude shaping (PAS) is being investigated as a way to improve performance of fiber-optic communication systems ~\cite{Dar2014a,Geller2016,Amari2019,Cho2019}. PAS allows to finely adapt the information rate and has also been demonstrated to have improved tolerance to the fiber nonlinearity ~\cite{Fehenberger2016,Amari2019,Pilori2019,Fu2021}. In recent works ~\cite{Mello2018,Civelli2020,Borujeny2023,Civelli2023}, it was shown that the nonlinear shaping gain from probabilistic shaping is reduced in the presence of a carrier phase estimation (CPE) system. A CPE system is essential in practical systems as it is required to correct the phase noise arising from non-ideal lasers. CPE also appears to aid in nonlinearity compensation as the fiber nonlinearity manifests partly as a phase noise on the received symbols ~\cite{Mecozzi2012}. Hence, it is beneficial to design new CPE systems or modify existing CPE systems to aid in nonlinearity mitigation ~\cite{Rosa2020}.

The blind phase search (BPS) algorithm ~\cite{Pfau2009} is a standard feed-forward CPE algorithm used in fiber-optic communication systems ~\cite{Mello2018}. By design, the BPS algorithm does not take into account the channel parameters and transmitted symbol probabilities. While this makes the BPS algorithm versatile and easy to implement, its performance is sub-optimal for probabilistically shaped constellations ~\cite{Yan2021}. The maximum a posteriori (MAP) CPE for the Wiener phase noise channel was investigated in ~\cite{RiosMuller2017}. The BPS algorithm is essentially a simplification of the MAP CPE. It was shown that the MAP CPE has a superior residual phase noise performance compared to the BPS algorithm. However, a Monte Carlo integration method was used in ~\cite{RiosMuller2017} to implement the MAP CPE which has an enormous computational cost, rendering it impractical.

In this paper, we demonstrate that an approximate MAP CPE for the Wiener phase noise channel can be implemented using belief propagation (BP) at a reasonable cost. We show that it has improved bit-wise mutual information (BMI) performance compared to the BPS algorithm. The improvement however cannot always be justified by the increased computational cost. Hence as a low-complexity alternative we explore machine learning based improvements to the BPS algorithm and show that the performance of the BPS algorithm can be improved considerably with minimal changes.
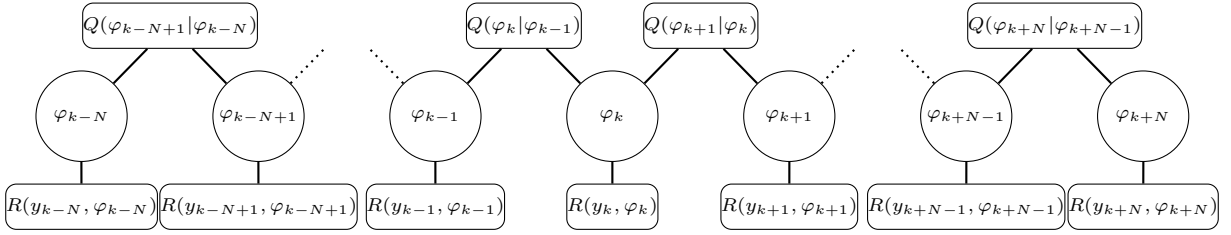
\begin{figure*}
\centering
    \begin{tikzpicture}[scale=1.2]
		\tikzstyle{variable node}=[circle, draw=black,  inner sep=0pt, minimum size=12mm]
		\tikzstyle{factor node}=[rectangle,rounded corners, draw=black,  inner sep=0pt, minimum size=6mm]
            \tikzstyle{middle node}=[rectangle, draw=black, fill=red!20, inner sep=0pt, minimum size=6mm]
            \tikzstyle{equation node}=[rectangle,  inner sep=0pt, minimum size=6mm]
            \tikzstyle{invisible node}=[rectangle,  inner sep=0pt, minimum size=6mm]
            \def\tsize{\scriptsize}
            \def\val{0.97}
            \node[variable node] (pn1) at (-6*\val,0) []{\tsize$\varphi_{k-N}$};
            \node[variable node] (pn2) at (-4*\val,0) []{\tsize$\varphi_{k-N+1}$};
            \node[variable node] (pn3) at (-2*\val,0) []{\tsize$\varphi_{k-1}$};
            \node[variable node] (pn4) at (-0,0) []{\tsize$\varphi_{k}$};
            \node[variable node] (pn5) at (2*\val,0) []{\tsize$\varphi_{k+1}$};
            \node[variable node] (pn6) at (4*\val,0) []{\tsize$\varphi_{k+N-1}$};
            \node[variable node] (pn7) at (6*\val,0) []{\tsize$\varphi_{k+N}$};

            \node[factor node] (fn1) at (-5*\val,1) []{\tsize$Q(\varphi_{k-N+1}|\varphi_{k-N})$};
            \node[factor node] (fn2) at (-1*\val,1) []{\tsize$Q(\varphi_{k}|\varphi_{k-1})$};
            \node[factor node] (fn3) at (1*\val,1) []{\tsize$Q(\varphi_{k+1}|\varphi_{k})$};
            \node[factor node] (fn4) at (5*\val,1) []{\tsize$Q(\varphi_{k+N}|\varphi_{k+N-1})$};

            \node[factor node] (rn1) at (-6*\val,-1) []{\tsize $R(y_{k-N},\varphi_{k-N})$};
            \node[factor node] (rn2) at (-4*\val,-1) []{\tsize $R(y_{k-N+1},\varphi_{k-N+1})$};
            \node[factor node] (rn3) at (-2*\val,-1) []{\tsize $R(y_{k-1},\varphi_{k-1})$};
            \node[factor node] (rn4) at (-0*\val,-1) []{\tsize$R(y_{k},\varphi_{k})$};
            \node[factor node] (rn5) at (2*\val,-1) []{\tsize$R(y_{k+1},\varphi_{k+1})$};
            \node[factor node] (rn6) at (4*\val,-1) []{\tsize$R(y_{k+N-1},\varphi_{k+N-1})$};
            \node[factor node] (rn7) at (6*\val,-1) []{\tsize$R(y_{k+N},\varphi_{k+N})$};

            \foreach \x in {1,2,...,7}
                \draw [-,thick] (pn\x) -- (rn\x);

            \draw [-,thick] (pn1) -- (fn1);
            \draw [-,thick] (pn2) -- (fn1);
            \draw [-,thick] (pn3) -- (fn2);
            \draw [-,thick] (pn4) -- (fn2);
            \draw [-,thick] (pn4) -- (fn3);
            \draw [-,thick] (pn5) -- (fn3);
            \draw [-,thick] (pn6) -- (fn4);
            \draw [-,thick] (pn7) -- (fn4);

            \node[invisible node] (in1) at (-3*\val,1) []{};
            \node[invisible node] (in2) at (3*\val,1) []{};

            \draw [dotted,thick] (pn2) -- (in1);
            \draw [dotted,thick] (pn3) -- (in1);
            \draw [dotted,thick] (pn5) -- (in2);
            \draw [dotted,thick] (pn6) -- (in2);

	\end{tikzpicture}
 \caption{Factor graph of the product term in Eq. \ref{eq:MAP_approx}}
 \label{fig:Factor_graph}
\end{figure*}

\section{MAP CPE}
We work with the Wiener phase noise channel which has the following discrete model
\begin{equation}
    y_k=x_k\mathrm{e}^{\mathrm{j}\varphi_k}+n_k,
    \label{eq:channel}
\end{equation}
where ${x_k\in \mathcal{X}\subset\mathbb{C}}$ are transmitted symbols with probability distribution $P(x)$ and ${y_k\in \mathbb{C}}$ are the received symbols. The symbol $x_k$ at time instant $k$ is affected by phase noise modeled as ${\varphi_k=\varphi_{k-1}+\theta_k}$, ${\theta_k \sim \mathcal{N}(0,\sigma_\theta^2)}$ and circular additive-white Gaussian noise (AWGN) ${n_k\sim \mathcal{C}\mathcal{N}(0,\sigma_n^2)}$. Given $2N+1$ received symbols $\bm{y}=[y_{k-N},\dots,y_k,\dots,y_{k+N}]$, the MAP phase estimate ~\cite{RiosMuller2017} is given by ${\hat{\varphi}_k=\argmax_{\varphi_k} P(\varphi_k|\bm{y})}$.
\vspace{-1em}
\begin{equation}   
\begin{aligned}
    P(\varphi_k|\bm{y})=\int \cdots \int \prod_{i=k-N}^{k+N}R(y_i,\varphi_i)Q(\varphi_i|\varphi_{i-1})\\
    \qquad \qquad \qquad \mathrm{d}\varphi_{k-N}\cdots \mathrm{d}\varphi_{k-1}\mathrm{d}\varphi_{k+1}\cdots \mathrm{d}\varphi_{k+N}
\end{aligned}
\label{eq:MAP}
\end{equation}

where $R(y_i,\varphi_i)=\sum_{x\in \mathcal{X}} P(y_i|x,\varphi_i) P(x) $ with $P(y_i|x,\varphi_i)\propto\mathrm{exp}\left(-\frac{|y_i-x\mathrm{e}^{\mathrm{j}\varphi_i}|^2}{2\sigma^2_n} \right)$ due to the AWGN and $Q(\varphi_i|\varphi_{i-1})\propto\mathrm{exp}\left(\frac{-(\varphi_i-\varphi_{i-1})^2}{2\sigma^2_{\theta}}\right)$ due to the phase noise.

For numerical implementation, we assume that the $\varphi_k$ can only take on a finite set of phase values in  $\bm{\phi}=\{\phi_1,\phi_2,\dots,\phi_M\}$. Then the approximate MAP estimator is given by
\begin{equation}
\begin{split}
    \hat{\varphi}_k&=\argmax_{\varphi_k \in \bm{\phi}} \sum_{\varphi_{k-N}\in \bm{\phi}} \cdots\sum_{\varphi_{k-1}\in \bm{\phi}}\sum_{\varphi_{k+1}\in \bm{\phi}} \cdots \\ 
     &\sum_{\varphi_{k+N}\in \bm{\phi}} \prod_{i=k-N}^{k+N}R(y_i,\varphi_i)Q(\varphi_i|\varphi_{i-1}).
\end{split}
\label{eq:MAP_approx}
\end{equation}

Assuming phase noise remains constant in the window of $2N+1$ symbols and equiprobable transmit symbols, considering only the constellation symbol that provides the largest contribution in $P(y_i|x,\varphi_i)$ yields the BPS algorithm:
\begin{align}
    \hat{\varphi}_k&=\argmax_{\varphi\in \bm{\phi}} \prod_{i=k-N}^{k+N}\mathrm{e}^{-\frac{|y_i-\hat{x}_n\mathrm{e}^{\mathrm{j}\varphi}|^2}{2\sigma^2_n}} \nonumber \\
    &=\argmin_{\varphi\in \bm{\phi}} \sum_{i=k-N}^{k+N}    |y_i-\hat{x}_n\mathrm{e}^{\mathrm{j}\varphi}|^2 ,\label{eq:BPS}
\end{align}
where $\hat{x}_n=\argmin_{x\in\mathcal{X}} |y_n-x\mathrm{e}^{\mathrm{j}\varphi}|^2$.

\section{Implementation of MAP CPE using BP}
The MAP phase estimation problem in (\ref{eq:MAP_approx}) is a marginalization problem which can be implemented efficiently using belief propagation. We can represent the product term in (\ref{eq:MAP_approx}) as the factor graph in Fig. \ref{fig:Factor_graph}. As the graph in Fig. \ref{fig:Factor_graph} is a tree, the marginalization problem can be solved exactly by applying the sum-product algorithm on the respective graph ~\cite{Kschischang2001}. The specific algorithm for obtaining the phase estimate $\hat{\varphi}_k$ in (\ref{eq:MAP_approx}) is given in Alg. \ref{alg:MAPFG}.

\begin{algorithm}
\caption{Approximate MAP phase estimation using BP}\label{alg:MAPFG}
\begin{algorithmic}
\Require $\bm{y}$, $\bm{\phi}$, $R(y_i,\varphi_i)$, $Q(\varphi_i|\varphi_{i-1})$
\State $\bm{Q} \gets Q(\bm{\phi}|\bm{\phi})\in\mathbb{R}^{M\times M}$
\State $\bm{m_-} \gets (1,1,\dots,1)^\intercal\in\mathbb{R}^M$
\State $\bm{m_+} \gets (1,1,\dots,1)^\intercal\in\mathbb{R}^M$
\For{$i=k-N$ to $k-1$}
    \State $\bm{m_-} \gets \bm{Q} \left(\bm{m_-}\odot R(y_i,\bm{\phi})\right)$
    \LineComment {$\odot$ Hadamard product}
\EndFor
\For{$i=k+N$ to $k+1$}
    \State $\bm{m_+} \gets \bm{Q} \left(\bm{m_+}\odot R(y_i,\bm{\phi})\right)$
\EndFor
\State \textbf{return} $\hat{\varphi}_k \gets \argmax_{\varphi \in \bm{\phi}}  \bm{m_-}\odot R(y_k,\bm{\phi})\odot \bm{m_+} $
\end{algorithmic}
\end{algorithm}

\section{Simulation results}
We perform numerical simulations on the Wiener phase noise channel (\ref{eq:channel}) with Maxwell-Boltzmann shaped 64-QAM constellations for varying signal-to-noise ratios (SNRs) and $\sigma^2_{\theta}$ values. We set the test angles as $\phi_i=-\pi/n + \frac{(i-1)2\pi}{nM}$ for $i=1,\dots,M$ with $n$ equal to the degree of rotational symmetry of the constellation. For QAM constellations, we have $n=4$. To account for phase wrapping, we use the wrapped normal distribution for the phase noise
\begin{equation*}
       Q(\varphi_i|\varphi_{i-1})\propto\sum\limits_{\mathclap{r=-\infty}}^{\mathclap{r=\infty}}\mathrm{exp}\mleft(\frac{-(\varphi_i-\varphi_{i-1}+2\pi r/n)^2}{2\sigma^2_{\theta}}\mright).
       \label{eq:wrapped_normal_distribution}
\end{equation*}
A mismatched circular Gaussian demapper is used with optimized noise variance. We use the BMI calculated using the method from ~\cite{Fehenberger2016} to compare the performance of the new estimator (MAP) with the BPS algorithm and the algorithm with constant phase noise assumption from ~\cite{RiosMuller2017} (CPN). Phase unwrapping is applied to the estimated phase sequences and a fully data-aided cycle-slip compensation is used. We simulate sequences of $2^{15}$ symbols and report the median BMI value over $100$ realizations. 
\begin{figure}[t]
  \centering
  \includestandalone[width=\columnwidth]{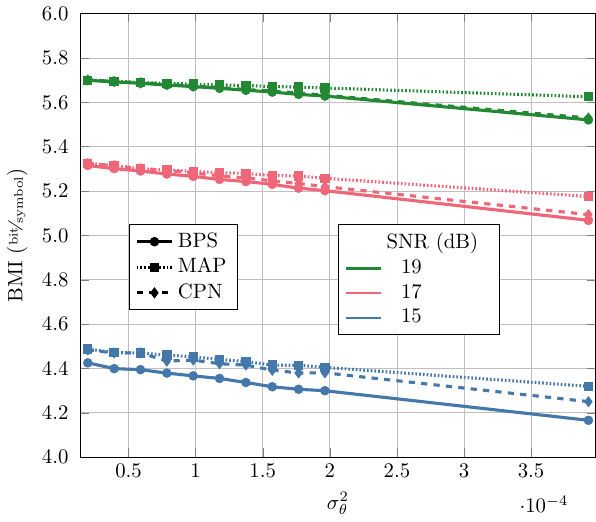}  
  \caption{Performance comparison between CPE algorithms with $N=32$ and $M=60$ for varying channel parameters.}
  \label{fig:performance_plot_60tp}
\end{figure}
In Fig. \ref{fig:performance_plot_60tp}, we can see the results for half-window size $N=32$ and $M=60$ test phases. Of the three estimators in Fig. \ref{fig:performance_plot_60tp}, the MAP estimator has the best performance. The improvement at higher $\sigma^2_{\theta}$ values can be attributed to the use of a model for the phase noise. While the gains of the new estimator are clear in Fig. \ref{fig:performance_plot_60tp}, they are diminished when a more practical number of $M=15$ test phases is used, as shown in Fig. \ref{fig:performance_plot_15tp}. For lower number of test phases, the approximation (\ref{eq:MAP_approx}) deviates significantly from the MAP estimator (\ref{eq:MAP}) which explains the reduced improvement in Fig. \ref{fig:performance_plot_15tp}. In this scenario the additional cost of the MAP estimator cannot be justified by the improvement in BMI. We hence explore machine learning based improvements to the BPS algorithm.

\section{Improved BPS}
We start by modifying the differentiable BPS algorithm from ~\cite{Rode2023} as follows: 
\begin{align}
    d_{i,m} &= \min_{x\in\mathcal{X}} |y_l-x\mathrm{e}^{\mathrm{j}\phi_m}|^2 \nonumber\\
    &\forall \: i\in\{k-N,\dots,k+N\},\: \forall \: m\in\{1,\dots,M\}  \nonumber \\
    D_m &= \sum_{i=k-N}^{k+N} w_i d_{i,m}  \nonumber \\
    \hat{\varphi}_k&=\frac{\arg
    \left(\mathrm{e}^{\mathrm{j}\bm{\phi}n} \cdot \mathrm{softmin}_t(\bm{D})\right)}{n},\label{eq:BPS_improved}
\end{align}    
An arbitrary normalization $\sum_{i=k-N}^{k+N} w_i=1$ is chosen and the softmin with temperature $t$ is defined as
% \vspace{-1em}
\begin{equation}
    \mathrm{softmin}_t(x_i) = (\mathrm{softmin}_t(\bm{x}))_i=\frac{\mathrm{e}^{-\frac{x_i}{t}}}{\sum_j \mathrm{e}^{-\frac{x_j}{t}}}
    \label{eq:softmint}
\end{equation}
% \vspace{-1em}

\begin{figure}[t]
  \centering
  \includestandalone[width=\columnwidth]{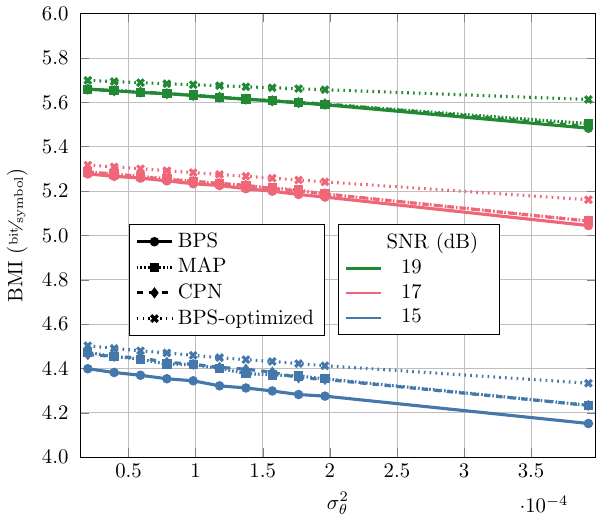}
  \caption{Performance comparison between CPE algorithms with $N=32$ and $M=15$ for varying channel parameters.}
  \label{fig:performance_plot_15tp}
\end{figure}

\begin{figure}[t]
  \centering
  \vspace{1em}
  \includestandalone[width=\columnwidth]{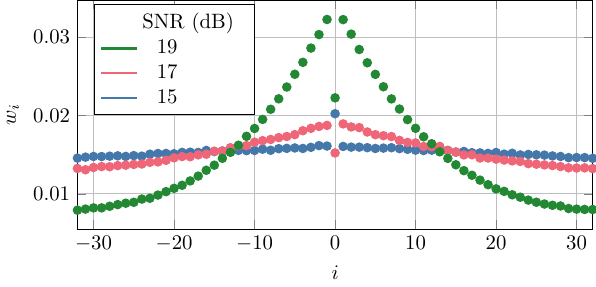}
  \caption{Weights $w_i$ learned by the BPS-optimized algorithm with $N=32$ and $M=15$ for $\sigma^2_{\theta}=\num{1.18e-4}$.}
  \label{fig:learned_weights}
\end{figure}

The first difference to ~\cite{Rode2023} is the use of weights~$w_i$ and the second difference is that the dot-product is taken between $\mathrm{softmin}_t(\bm{D})$ and $\mathrm{e}^{j\bm{\phi}n}$ rather than $\bm{\phi}n$. The second difference solves the performance degradation problem due to the phase discontinuity reported in ~\cite[Sec. VI.C]{Rode2023}. For uniform weights $w_i=1/(2N+1)$ and temperature $t\to 0$, we recover the standard BPS algorithm. Setting $N=32$ and $M=15$, for each value of SNR and of $\sigma^2_{\theta}$, we learn the weights $\bm{w}$ and temperature $t$ using an end-to-end optimization approach similar to the one used in ~\cite{Rode2023}. Training is performed over $100$ epochs using the Adam optimizer in PyTorch with a learning rate of $10^{-3}$. The number of batches is increased from $10$ to $100$ and the batch size is increased from $2^{12}$ to $2^{17}$ symbols over the $100$ epochs. We report the performance of the improved BPS algorithm (BPS-optimized) in Fig. \ref{fig:performance_plot_15tp}. The performance of the BPS-optimized algorithm with $M=15$ matches the performance of the MAP algorithm with $M=60$, at a significantly lower computational cost. In Fig. \ref{fig:learned_weights} we show the weights $w_i$ learned by the BPS-optimized algorithm for $\sigma^2_{\theta}=\num{1.18e-4}$.

%-------------------------------------------------- Conclusions Section ———————————————————————————%

\section{Conclusions}
We firstly presented a BP based approximate MAP CPE and demonstrated its superior performance compared to the standard BPS algorithm, especially for higher values of phase noise variance $\sigma^2_{\theta}$. The performance gain is diminished when a small number of test angles are used. We then proposed improvements to the BPS algorithm using end-to-end learning. The improved BPS algorithm has performance similar to the MAP CPE at a lower computational cost. End-to-end learning may be used similarly over a differentiable model of the fiber-optic channel for optimization of the BPS algorithm to aid in non-linearity mitigation.

%-------------------------------------------------- Acknowledgements Section -------------------------------------------------------%

% \section{Acknowledgments}
% This work has received funding from the European Research Council (ERC) under the European Union's Horizon 2020 research and innovation programme (grant agreement No. 101001899). 

%-------------------------------------------------- Bibliography Section -------------------------------------------------------%
\balance
\clearpage
\printbibliography

@Article{Fehenberger2016,
  author    = {Fehenberger, Tobias and Alvarado, Alex and Bocherer, Georg and Hanik, Norbert},
  journal   = {Journal of Lightwave Technology},
  title     = {On Probabilistic Shaping of Quadrature Amplitude Modulation for the Nonlinear Fiber Channel},
  year      = {2016},
  month     = nov,
  number    = {21},
  pages     = {5063--5073},
  volume    = {34},
  doi       = {10.1109/jlt.2016.2594271},
  publisher = {Institute of Electrical and Electronics Engineers ({IEEE})},
}

@InProceedings{Civelli2020,
  author    = {Civelli, Stella and Forestieri, Enrico and Secondini, Marco},
  booktitle = {2020 European Conference on Optical Communications ({ECOC})},
  title     = {Interplay of Probabilistic Shaping and Carrier Phase Recovery for Nonlinearity Mitigation},
  year      = {2020},
  month     = dec,
  publisher = {{IEEE}},
  doi       = {10.1109/ecoc48923.2020.9333212},
}

@Article{Pfau2009,
  author    = {Pfau, Timo and Hoffmann, Sebastian and Noe, Reinhold},
  journal   = {Journal of Lightwave Technology},
  title     = {Hardware-Efficient Coherent Digital Receiver Concept With Feedforward Carrier Recovery for $M$-{QAM} Constellations},
  year      = {2009},
  month     = apr,
  number    = {8},
  pages     = {989--999},
  volume    = {27},
  doi       = {10.1109/jlt.2008.2010511},
  publisher = {Institute of Electrical and Electronics Engineers ({IEEE})},
}

@Article{Fu2021,
  author    = {Fu, Mengfan and Liu, Qiaoya and Lun, Huazhi and Jiang, Hexun and Wu, Yiwen and Liu, Xiaomin and Yang, Zhiyuan and Yi, Lilin and Hu, Weisheng and Zhuge, Qunbi},
  journal   = {Journal of Lightwave Technology},
  title     = {Parallel Bisection-based Distribution Matching for Nonlinearity-tolerant Probabilistic Shaping in Coherent Optical Communication Systems},
  year      = {2021},
  month     = oct,
  number    = {20},
  pages     = {6459--6469},
  volume    = {39},
  doi       = {10.1109/jlt.2021.3102269},
  publisher = {Institute of Electrical and Electronics Engineers ({IEEE})},
}

@InProceedings{Rosa2020,
  author    = {di Rosa, Gabriele and Richter, Andre},
  booktitle = {2020 22nd International Conference on Transparent Optical Networks ({ICTON})},
  title     = {Achievable Mitigation of Nonlinear Phase Noise through Optimized Blind Carrier Phase Recovery},
  year      = {2020},
  month     = jul,
  publisher = {{IEEE}},
  doi       = {10.1109/icton51198.2020.9203148},
}

@InProceedings{Yan2021,
  author    = {Yan, Qifeng and Xia, Yu and Zhou, Yujun and Zhang, Peishan and Hong, Xuezhi},
  booktitle = {2021 Asia Communications and Photonics Conference (ACP)},
  title     = {Low-Complexity Carrier Phase Estimation for Probabilistically Shaped {MQAM} Signals based on {Kullback-Leibler} Divergence Analysis with Squared Symbols},
  year      = {2021},
  pages     = {1-3},
  publisher = {Optica Publishing Group},
  doi       = {10.1364/ACPC.2021.M4I.4},
}

@InProceedings{RiosMuller2017,
  author    = {Rios-Muller, Rafael and Bitachon, Bertold Ian},
  booktitle = {2017 European Conference on Optical Communication ({ECOC})},
  title     = {Maximum Likelihood Carrier Phase Estimation Based on {Monte Carlo} Integration},
  year      = {2017},
  month     = sep,
  publisher = {{IEEE}},
  doi       = {10.1109/ecoc.2017.8345920},
}

@Article{Civelli2023,
  author    = {Stella Civelli and Emanuele Parente and Enrico Forestieri and Marco Secondini},
  journal   = {Journal of Lightwave Technology},
  title     = {On the Nonlinear Shaping Gain with Probabilistic Shaping and Carrier Phase Recovery},
  year      = {2023},
  pages     = {1--12},
  doi       = {10.1109/jlt.2023.3241449},
  publisher = {Institute of Electrical and Electronics Engineers ({IEEE})},
}

@Article{Rode2023,
  author    = {Rode, Andrej and Geiger, Benedikt and Chimmalgi, Shrinivas and Schmalen, Laurent},
  journal   = {Journal of Lightwave Technology},
  title     = {End-to-end Optimization of Constellation Shaping for {Wiener} Phase Noise Channels with a Differentiable Blind Phase Search},
  year      = {2023},
  pages     = {1--11},
  doi       = {10.1109/jlt.2023.3265308},
  publisher = {Institute of Electrical and Electronics Engineers ({IEEE})},
}

@Article{Kschischang2001,
  author    = {Kschischang, F. R. and Frey, B. J. and Loeliger, H.-A.},
  journal   = {{IEEE} Transactions on Information Theory},
  title     = {Factor graphs and the sum-product algorithm},
  year      = {2001},
  number    = {2},
  pages     = {498--519},
  volume    = {47},
  doi       = {10.1109/18.910572},
  publisher = {Institute of Electrical and Electronics Engineers ({IEEE})},
}

@Article{Borujeny2023,
  author    = {Reza Rafie Borujeny and Frank R. Kschischang},
  journal   = {Journal of Lightwave Technology},
  title     = {Why Constant-Composition Codes Reduce Nonlinear Interference Noise},
  year      = {2023},
  pages     = {1--8},
  doi       = {10.1109/jlt.2023.3244927},
  publisher = {Institute of Electrical and Electronics Engineers ({IEEE})},
}

@Article{Amari2019,
  author    = {Amari, Abdelkerim and Goossens, Sebastiaan and Gultekin, Yunus Can and Vassilieva, Olga and Kim, Inwoong and Ikeuchi, Tadashi and Okonkwo, Chigo M. and Willems, Frans M. J. and Alvarado, Alex},
  journal   = {Journal of Lightwave Technology},
  title     = {Introducing Enumerative Sphere Shaping for Optical Communication Systems With Short Blocklengths},
  year      = {2019},
  month     = dec,
  number    = {23},
  pages     = {5926--5936},
  volume    = {37},
  doi       = {10.1109/jlt.2019.2943938},
  publisher = {Institute of Electrical and Electronics Engineers ({IEEE})},
}

@Article{Geller2016,
  author    = {Geller, Omri and Dar, Ronen and Feder, Meir and Shtaif, Mark},
  journal   = {Journal of Lightwave Technology},
  title     = {A Shaping Algorithm for Mitigating Inter-Channel Nonlinear Phase-Noise in Nonlinear Fiber Systems},
  year      = {2016},
  month     = aug,
  number    = {16},
  pages     = {3884--3889},
  volume    = {34},
  doi       = {10.1109/jlt.2016.2575400},
  publisher = {Institute of Electrical and Electronics Engineers ({IEEE})},
}

@Article{Mecozzi2012,
  author    = {Mecozzi, Antonio and Essiambre, Ren{\'{e}}-Jean},
  journal   = {Journal of Lightwave Technology},
  title     = {Nonlinear {S}hannon Limit in Pseudolinear Coherent Systems},
  year      = {2012},
  month     = jun,
  number    = {12},
  pages     = {2011--2024},
  volume    = {30},
  doi       = {10.1109/jlt.2012.2190582},
  publisher = {Institute of Electrical and Electronics Engineers ({IEEE})},
}

@InProceedings{Dar2014a,
  author    = {Dar, Ronen and Feder, Meir and Mecozzi, Antonio and Shtaif, Mark},
  booktitle = {2014 {IEEE} International Symposium on Information Theory},
  title     = {On shaping gain in the nonlinear fiber-optic channel},
  year      = {2014},
  month     = jun,
  publisher = {{IEEE}},
  doi       = {10.1109/isit.2014.6875343},
}

@Article{Pilori2019,
  author    = {Pilori, Dario and Nespola, Antonino and Forghieri, Fabrizio and Bosco, Gabriella},
  journal   = {Journal of Lightwave Technology},
  title     = {Non-Linear Phase Noise Mitigation Over Systems Using Constellation Shaping},
  year      = {2019},
  month     = jul,
  number    = {14},
  pages     = {3475--3482},
  volume    = {37},
  doi       = {10.1109/jlt.2019.2917308},
  publisher = {Institute of Electrical and Electronics Engineers ({IEEE})},
}

@Article{Cho2019,
  author    = {Cho, Junho and Winzer, Peter J.},
  journal   = {Journal of Lightwave Technology},
  title     = {Probabilistic Constellation Shaping for Optical Fiber Communications},
  year      = {2019},
  month     = mar,
  number    = {6},
  pages     = {1590--1607},
  volume    = {37},
  doi       = {10.1109/jlt.2019.2898855},
  publisher = {Institute of Electrical and Electronics Engineers ({IEEE})},
}

@Article{Mello2018,
  author    = {Mello, Darli A. A. and Barbosa, Fabio Aparecido and Reis, Jacklyn Dias},
  journal   = {Journal of Lightwave Technology},
  title     = {Interplay of Probabilistic Shaping and the Blind Phase Search Algorithm},
  year      = {2018},
  month     = nov,
  number    = {22},
  pages     = {5096--5105},
  volume    = {36},
  doi       = {10.1109/jlt.2018.2869245},
  publisher = {Institute of Electrical and Electronics Engineers ({IEEE})},
}

\vspace{-4mm}

%%%%%%%%%%%%%%%%%%%%%%%%%%%%%%%%%%%%%%%%%%%%%
%---------------------------------------------- End of Document -----------------------------------------------%
\end{document}